\documentclass[sigconf,nonacm]{acmart}

\acmBadgeR{}
\acmBadgeL{}
\renewcommand\footnotetextcopyrightpermission[1]{} 
\usepackage{graphicx}
\usepackage{booktabs} 

\usepackage{hyperref}

\usepackage[utf8]{inputenc}
\usepackage{caption}
\usepackage{multirow}
\usepackage{xcolor}
\usepackage{graphicx}
\usepackage{epstopdf}
\usepackage{url}
\usepackage{comment}
\usepackage{graphicx}
\usepackage{caption}
\usepackage{subcaption}
\usepackage{tabularx}

\usepackage{enumitem}
\setlist[itemize]{align=parleft,left=0pt..1em}

\AtBeginDocument{%
  \providecommand\BibTeX{{%
  \normalfont 
  \kern-0.5em{\scshape i\kern-0.25em b}\kern-0.8em\TeX}
  }}

\begin{document}

\title{STRAPSim: A Portfolio Similarity Metric for ETF Alignment and Portfolio Trades}



    
\author{Mingshu Li}
\email{mingshu.li@blackrock.com}
\affiliation{%
  \institution{BlackRock, Inc.}
  \city{Atlanta}
  \state{GA}
  \country{USA}
}
\author{Dhruv Desai}
\email{dhruv.desai1@blackrock.com}
\affiliation{%
  \institution{BlackRock, Inc.}
 \city{New York}
  \state{NY}
  \country{USA}
}

\author{Jerinsh Jeyapaulraj}
\email{jerinsh244@gmail.com }
\affiliation{%
  \institution{BlackRock, Inc.}
 \city{New York}
  \state{NY}
  \country{USA}
}

\author{Philip Sommer}
\email{philip.sommer@blackrock.com}
\affiliation{%
  \institution{BlackRock, Inc.}
 \city{New York}
  \state{NY}
  \country{USA}
}

\author{Riya Jain}
\email{riya.jain@blackrock.com}
\affiliation{%
  \institution{BlackRock, Inc.}
 \city{Gurugram}
  \state{HR}
  \country{India}
}

\author{Peter Chu}
\email{peter.chu@blackrock.com}
\affiliation{%
  \institution{BlackRock, Inc.}
 \city{New York}
  \state{NY}
  \country{USA}
}

\author{Dhagash Mehta}
\email{dhagash.mehta@blackrock.com}
\affiliation{%
  \institution{BlackRock, Inc.}
  \city{New York, NY}
  \country{USA}
}

\renewcommand{\shortauthors}{Li et al.}

\begin{abstract}
Accurately measuring portfolio similarity is critical for a wide range of financial applications, including Exchange-traded Fund (ETF) recommendation, portfolio trading, and risk alignment. Existing similarity measures often rely on exact asset overlap or static distance metrics, which fail to capture similarities among the constituents (e.g., securities within the portfolio) as well as nuanced relationships between partially overlapping portfolios with heterogeneous weights. We introduce STRAPSim (\textbf{S}emantic, \textbf{T}wo-level, \textbf{R}esidual-\textbf{A}ware \textbf{P}ortfolio \textbf{Sim}ilarity), a novel method that computes portfolio similarity by matching constituents based on semantic similarity, weighting them according to their portfolio share, and aggregating results via residual-aware greedy alignment. We benchmark our approach against Jaccard, weighted Jaccard, as well as BERTScore-inspired variants across public classification, regression, and recommendation tasks, as well as on corporate bond ETF datasets. Empirical results show that our method consistently outperforms baselines in predictive accuracy and ranking alignment, achieving the highest Spearman correlation with return-based similarity. By leveraging constituent-aware matching and dynamic reweighting, portfolio similarity offers a scalable, interpretable framework for comparing structured asset baskets, demonstrating its utility in ETF benchmarking, portfolio construction, and systematic execution.
\end{abstract}
\keywords{Exchange-Traded Funds, Similarity Learning, Bond Similarity, ETF constituents, Portfolio Trades}

\maketitle

\graphicspath{ {plots/} }

\section{Introduction}
Exchange-Traded Funds (ETFs) and mutual funds are essential vehicles for diversified investment, offering investors access to broad asset classes through a single product. The ability to assess similarity between these portfolios is critical for a variety of financial decision-making tasks. These include fund recommendation, portfolio rebalancing, in-kind transfers, tax-loss harvesting, and liquidity analysis. For both institutional and retail investors, identifying portfolios with similar exposures enables improvements in asset allocation, risk management, and transaction efficiency.

For instance, recommending funds with similar risk-return profiles or portfolio compositions to those already held can improve diversification, reduce costs, and optimize exposure. Institutional workflows, such as in-kind transfers, where ETF shares are exchanged for a corresponding basket of securities, also rely on accurately assessing the overlap between asset holdings. Here, portfolio similarity directly impacts pricing confidence, hedging efficiency, and trade execution.

Despite its importance, portfolio similarity remains difficult to define and measure precisely. Common industry standards like Morningstar classifications rely on expert-driven taxonomies that group funds into broad categories \cite{blake2000morningstar,morningstarcategorization}, offering limited granularity and responsiveness to strategy shifts or subtle differences in asset composition \cite{herbas2023should}. Data-driven methods based on return correlations or aggregate features (e.g., beta, sector exposure, yield) also fall short, especially for less liquid assets like corporate bond ETFs, which may fail to capture partial overlaps or weight differences between portfolios designed to achieve similar objectives.

A growing body of work seeks to address these shortcomings using clustering methods \cite{marathe1999categorizing,desai2021robustness}, distance metric learning \cite{xing2002distance}, random forest proximities \cite{jeyapaulraj2022supervised}, or graph-based learning \cite{satone2021fund2vec}. However, these approaches either lack interpretability or fail to account for constituent-level similarities and weight distributions simultaneously, limiting their utility in high-stakes financial workflows.

Parallel to these modeling challenges, the rise of portfolio trading has transformed execution strategies in institutional finance. Portfolio trading involves executing a basket of bonds as a single, all‑or‑none transaction with one or more market‑makers, rather than trading each bond individually\cite{palleja2024portfolio,meli2023portfolio,madhavan2022trading}. This method enhances liquidity, price efficiency, and execution speed, particularly in less liquid credit markets. Its adoption has grown rapidly in recent years, and empirical studies have shown significant reductions in transaction costs, for example, over 40\% reduction for illiquid bonds, and improved access to institutional-scale credit market liquidity \cite{meli2023portfolio}. In the investment‑grade (IG) credit markets, on the other hand, portfolio trades constitute approximately 20\% of the U.S. market and 11\% of the European market in 2024\footnote{\url{https://www.ft.com/content/5e9f42b5-cb14-45fc-9804-279c28fd8418}}\footnote{\url{https://investor.marketaxess.com/news/news-details/2025/MarketAxess-Announces-Trading-Volume-Statistics-for-March-and-First-Quarter-2025/default.aspx}}. In the U.S., a portfolio trade occurs roughly every seven minutes, indicating its growing prominence as a preferred execution method for large and complex transactions.

Despite its growth, portfolio trading still faces challenges in aligning custom trade baskets with benchmark ETFs. Heuristic approaches, such as ticker overlap or sector tagging, lack precision and scalability, especially in credit markets with sparse pricing. Existing metrics often miss structural alignment when portfolios differ in holdings but share similar exposures. This underscores the need for a scalable, constituent-level similarity metric that captures both composition and weight distribution.

In this work, we propose a novel portfolio similarity metric, called STRAPSim (\textbf{S}emantic, \textbf{T}wo-level, \textbf{R}esidual-\textbf{A}ware \textbf{P}ortfolio \textbf{Sim}ilarity), that addresses the above limitations by explicitly incorporating both (1) constituent-level similarities and (2) weight distributions of the assets in individual portfolios under consideration. Unlike traditional set-based methods like the Jaccard index, our method dynamically reweights portfolio constituents as they are greedily matched, preserving the semantics of partial overlap while adapting to proportional differences.

In institutional fixed-income trading, when a portfolio trade closely resembles a benchmark ETF, dealers can price it efficiently and execute more quickly. However, they often rely on manual and heuristic methods to identify which ETF, if any, most closely matches a given portfolio trade. STRAPSim can enable a systematic approach by quantifying constituent-level similarity with weight awareness, allowing for more accurate basket matching and pricing. It can also support ETF basket creation and portfolio construction workflows, improving automation, transparency, and execution efficiency across institutional trading operations.

Beyond finance, the proposed methodology is broadly applicable to any domain requiring weighted set similarity, e.g., recommender systems, information retrieval, or structured document matching.

\subsection{Our Contribution and Previous Works}
A growing body of work has examined the problem of measuring similarity between financial portfolios, such as ETFs and mutual funds, due to its relevance in fund classification, recommendation, and trade alignment. Traditional methods often rely on set-based distance metrics such as the Jaccard index \cite{jaccard1901etude,costa2021further}, which quantify exact overlaps in holdings but ignore both the similarity between non-identical constituents and the distribution of portfolio weights.

To overcome these limitations, data-driven techniques have been introduced, including unsupervised clustering using k-means, hierarchical clustering, and self-organizing maps to group funds by attributes such as coupon rate, yield-to-maturity, and credit rating \cite{marathe1999categorizing, baghai2005consistency, bagde2018comprehensive, desai2021robustness, castellanos2024can}. Dimensionality reduction techniques like Principal Component Analysis (PCA) and Singular Value Decomposition (SVD) are commonly used to handle high-dimensional feature spaces \cite{wu2020similar}, and supervised models, such as random forests, have been applied to learn similarity or detect outlier funds based on proximity metrics \cite{jeyapaulraj2022supervised, desai2023quantifying}. Deep learning methods have also been employed to capture nonlinear dependencies between fund attributes \cite{kaya2019deep}, while graph-based approaches such as Fund2Vec \cite{satone2021fund2vec} represent fund-constituent relationships as bipartite graphs to infer structural similarity.

However, most of these methods do not operate directly on constituent-level holding data with associated weights. Instead, they focus on high-level summary features or latent representations, limiting their ability to reflect the true structure of the portfolios. Fund2Vec \cite{satone2021fund2vec} is a notable exceptions in that it utilizes holding level data for portfolio similarity, but its focus is to directly compute embeddings of the portfolios using a neural network and computing similarity between them. This method indeed takes into account the weights of the holdings within each portfolio rather than an interpretable, weight-sensitive similarity scoring.

Another line of work comes from semantic similarity in NLP, particularly BERTScore \cite{zhang2019bertscore}, which computes the similarity between two texts by matching each word to its most similar counterpart using contextual embeddings. BERTScore treats sentences as bags of words, which can be analogized to portfolios as bags of assets.While BERTScore-inspired methods use greedy matching to compute semantic similarity between unordered elements, they ignore relative importance (which is analogous to asset weights) and do not incorporate residual-aware matching, making them unsuitable for portfolios. Furthermore, they do not update residual weights after each match, which means elements can be matched multiple times without accounting for consumption. 

In this work, we propose a constituent-similarity-aware, weight-sensitive and residual-aware portfolio similarity metric that directly addresses these limitations. Given two portfolios, each defined by a set of constituents and associated weights, we first compute pairwise similarity scores using a user-defined function. We then perform greedy bipartite matching, aggregating scores based on the minimum available weight in each matched pair. After each match, constituent weights are dynamically updated, ensuring that no asset contributes more than its actual weight. This residual-aware mechanism prevents overcounting and captures the true degree of overlap between portfolios.

Our method has three key advantages: it operates directly on constituent-level asset data and takes portfolio weights into account; it supports interpretable and greedy matching based on domain-relevant similarity functions; and it dynamically adjusts weights after each match, preserving residual structure and avoiding duplication. This weight and constituent similarity-aware portfolio similarity, in turn, can be used to match custom trade baskets to benchmark ETFs, enabling more accurate pricing, better hedging, and scalable execution in portfolio trades.

\section{Proposed Method and other Benchmark Methodologies}
This section introduces STRAPSim, and basics of other benchmark methodologies, including Jaccard index, weighted Jaccard index, and BertScore-inspired metric.

\subsection{STRAPSIm Methodology}
\begin{figure}
    \centering
    \includegraphics[width=0.45\linewidth]{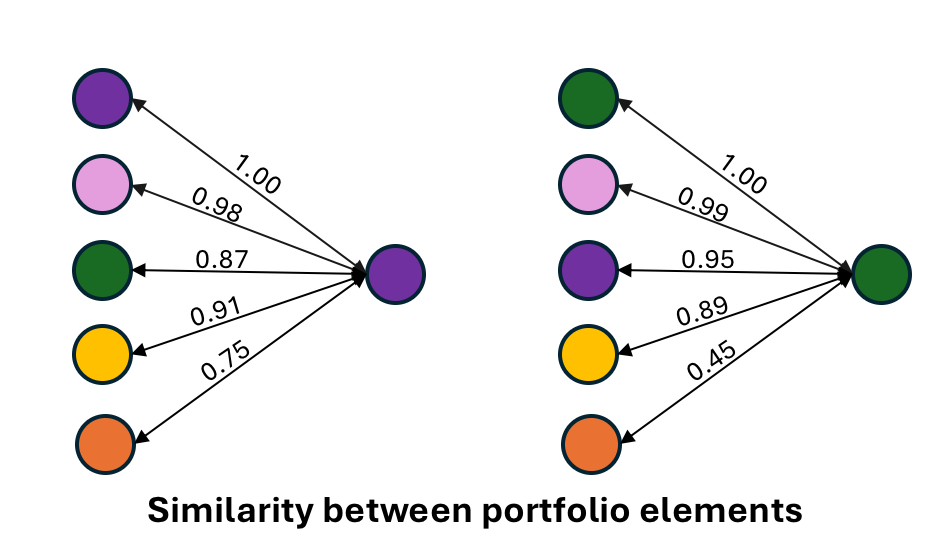}
    \includegraphics[width=0.45\linewidth]{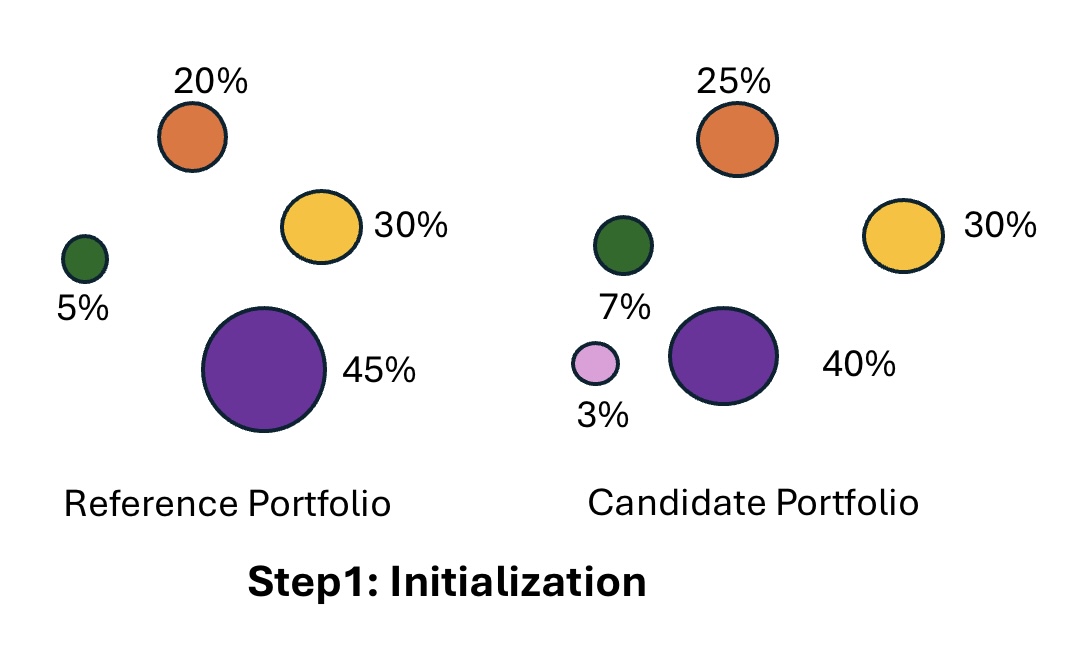}
    \includegraphics[width=0.45\linewidth]{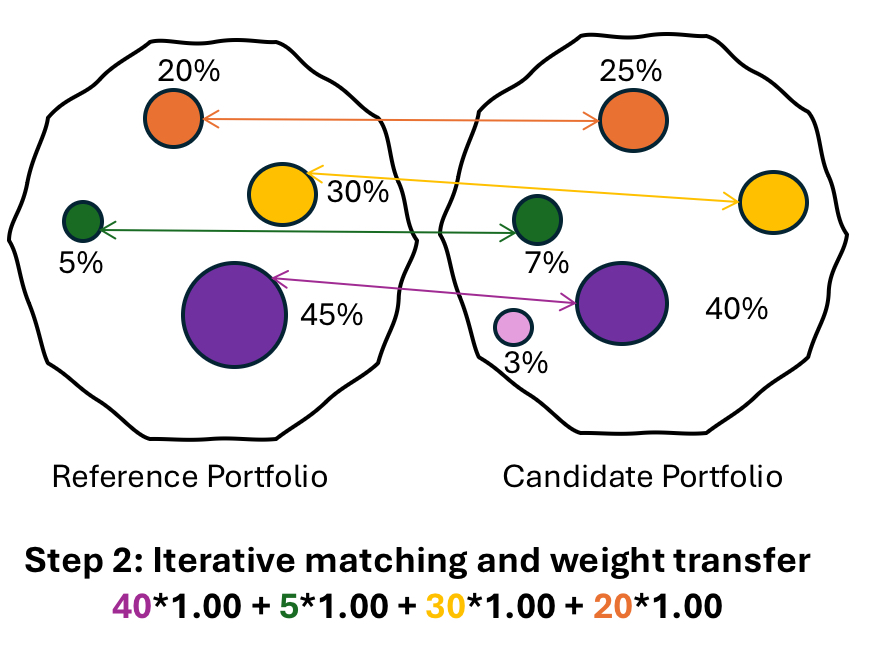}
    \includegraphics[width=0.45\linewidth]{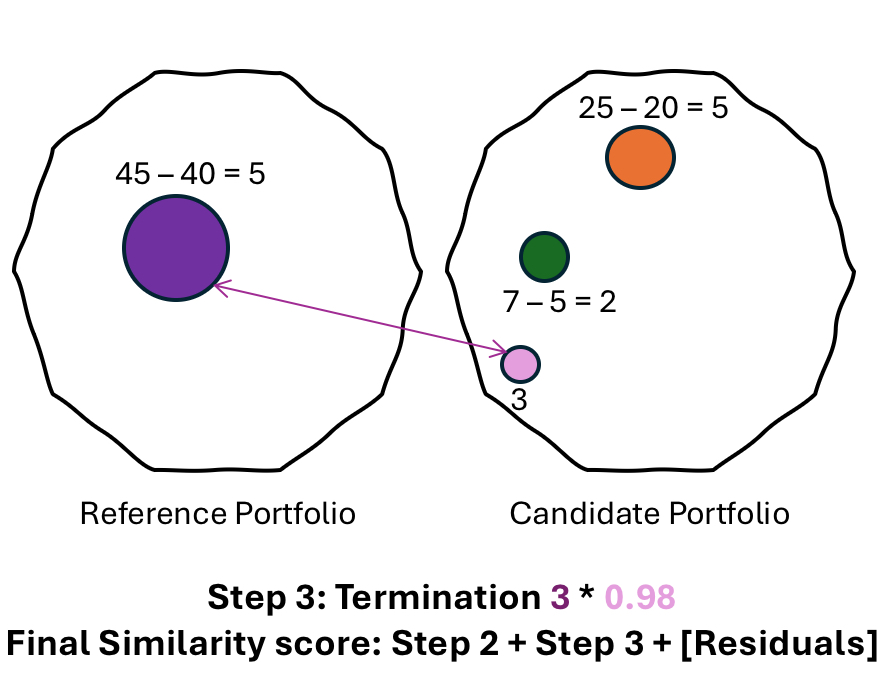}
    \caption {STRAPSim framework. The figure illustrates the similarity computation between elements of two portfolios using STRAPSim.
Left top: The similarity between portfolio elements is quantified; identical elements (e.g., purple–purple) have a similarity score of 1.00, while similar but non-identical elements (e.g., purple–pink) receive partial scores (e.g., 0.98).
Step 1 – Initialization: The composition of the reference portfolio (orange: 20\%, yellow: 30\%, green: 5\%, purple: 45\%) and the candidate portfolio (orange: 25\%, yellow: 40\%, green: 7\%, pink: 3\%) is established.
Step 2 – Iterative Matching and Weight Transfer: Elements from the reference portfolio are iteratively matched with similar elements in the candidate portfolio. Matches are based on the highest similarity scores, and the overlapping proportions are accumulated.
Step 3 – Termination and Residual Calculation: Unmatched or partially matched elements contribute to residuals. The final similarity score is computed as the sum of the weighted direct matches, indirect similar matches, and residual penalties.}
\Description[]{}
    \label{ref:STRAPSim_framework}
  \vspace{-5mm}
\end{figure}

The STRAPSim approach quantifies the similarity between two weighted sets by iteratively matching elements that exhibit the highest semantic alignment. Let $X$ be a reference set and $Y$ a candidate set, where each element in both sets is associated with a positive weight $w_{X}(i)$, $w_{Y}(j)$ and a semantic representation. For each element pair ($x_{i},y_{j}$), a pairwise similarity score $S_{ij}$ is computed using a chosen constituent-level similarity measure. The STRAPSim algorithm proceeds as follows (see also, Figure \ref{ref:STRAPSim_framework}):
\begin{enumerate}
    \item \textbf{Initialization:} All elements in $X$ and $Y$ are considered unmatched, with their full weights available.
    
    \item \textbf{Iterative Matching:} At each step, the pair of unmatched elements ($x_{i},y_{j}$) with the highest similarity score $S_{ij}$ is selected.
    
    \item \textbf{Weight Transfer:} The match contributes min($w_{X}(i)$, $w_{Y}(j)$ )$S_{ij}$ to the overall similarity score, where $w_{X}(i)$ and $w_{Y}(j)$  are the current available weights of $x_{i}$ and $y_{j}$,respectively
    
    \item \textbf{Weight Update:} The weights $w_{X}(i)$ and $w_{Y}(j)$  are updated by subtracting the matched amount. Elements are removed from further consideration once their weights reach zero.
    
    \item \textbf{Termination:} The process continues until no further significant weight remains to be matched.
    
\end{enumerate}
The final STRAPSim similarity score is the sum of all weighted pairwise similarities from the matching process. Additionally, a residual term captures the total unmatched weight remaining across both sets after the algorithm terminates.
Importantly, STRAPSim is agnostic to the specific constituent-level similarity function used to compute	$S_{ij}$; any appropriate semantic similarity measure can be integrated into the framework. In the following subsection, we introduce the specific similarity function employed in our experimental evaluation.

Eq.~(\ref{eq:bs})-(\ref{eq:bs_r}) presents the general mathematical formula for deriving STRAPSim that represents the most general conditions. 
\begin{subequations}
\begin{align}
  \label{eq:bs}
 STRAPSim(x,y) = \sum_{i,j=argsortS_{ij}}{S_{ij}\min({w_{X}^{(t)}(i),w_{Y}^{(t)}(j))}}, \\
 \intertext{with}
  \label{eq:bs_r}
  \begin{bmatrix}w_{X}^{(t+1)}(i) \\w_{Y}^{(t+1)}(j)\end{bmatrix}=
\begin{bmatrix}w_{X}^{(t)}(i) \\w_{Y}^{(t)}(j)\end{bmatrix}-\min{(w_{X}^{(t)}(i),w_{Y}^{(t)}(j))},
\end{align}
\end{subequations}
where $\text{argsort}(S)$ denotes the lexicographically largest permutation of the indices of the similarity matrix $S$, such that $S[\text{argsort}(S)]$ is sorted in descending order. The residual weights for each element after matching are given by $w_X^{(t)}(i)$ and $w_Y^{(t)}(j)$, respectively.

\subsubsection{Constituent-level Similarity Method}
Similarity estimation over structured data such as portfolios, customer records, or tabular entity sets, is a foundational task in machine learning, with applications ranging from clustering and recommendation to anomaly detection. Classical approaches include distance-based metrics (e.g., cosine, Euclidean, or Mahalanobis), kernel-based similarity, and recent advances in embedding-based or graph-based representations. These methods are typically unsupervised and rely on surface-level features or statistical transformations.

In contrast, supervised similarity learning seeks to learn a task-specific notion of similarity informed by labeled data. One effective method in this category is based on \textit{random forest proximities} \cite{breiman2001random}, which has been successfully applied to fund classification, entity matching, and outlier detection~\cite{jeyapaulraj2022supervised,saha2024machine}.

Random forest–based similarity is defined by the idea that instances which frequently fall into the same leaf nodes across an ensemble of decision trees are more likely to be similar. Formally, given a trained random forest classifier, the \textbf{proximity} between two samples $x_i$ and $x_j$ is calculated as shown in Eq.~\ref{eq:proximity}:

\begin{equation}\label{eq:proximity}
Proximity(x_{i},x_{j}) = \frac{1}{T}\sum_{t=1}^{T}\mathbb{I}[Leaf_{t}(x_{i}) = Leaf_{t}(x_{j})],
\end{equation}

where $T$ is the total number of trees in the forest, and $\mathbb{I}$ is the indicator function evaluating whether both instances land in the same leaf of tree $t$. Higher proximity score indicates greater similarity.

These proximities can then be used directly as similarity scores or transformed into distances for downstream tasks such as clustering or visualization. Notably, unlike classical similarity metrics, this approach implicitly captures feature interactions and non-linear boundaries present in the data, as learned during training. Additionally, the method inherits the interpretability and robustness of random forests \cite{lundberg2017unified,rosaler2024enhanced}, making it attractive for structured domains like finance and healthcare.

In the following, we use this random forest proximity based similarity method for the corporate bond ETFs experiment.

\subsection{Jaccard Index and Residual}
The Jaccard index \cite{jaccard1901etude} quantifies the similarity between finite sample sets by comparing the intersection and union of the sets. For two sets, $X$ and $Y$, the Jaccard index is defined as the ratio of the size of the intersection of $X$ and $Y$ to the size of their union (Eq.~\ref{eq:jaccard}). This metric is extensively utilized in various fields such as machine learning, recommendation systems, and natural language processing (NLP), where it is especially effective in assessing the presence and absence of elements within the sets. The Jaccard index values range from zero to one, with zero indicating no overlap and one indicating complete overlap between the sets. However, a significant limitation of the Jaccard index is that it does not account for the frequency of elements nor the relative sizes of the sets. The residual is defined as shown in Eq.~\ref{eq:jaccard}:
\begin{equation}\label{eq:jaccard}
J(X,Y) = \frac{|X\cap{Y}|}{|X\cup{Y}|}, \,
R_{J(X,Y)} = 1-\frac{|X\cap{Y}|}{|X\cup{Y}|},
\end{equation}
where $|X\cap{Y}|$ is the cardinality of the intersection of sets $X$ and $Y$, and $|X\cup{Y}|$ is the cardinality of the union of sets $X$ and $Y$.

\subsection{Weighted Jaccard Index and Residual}
The weighted Jaccard index \cite{costa2021further} extends the traditional approach by incorporating the weight of each element when computing the intersection and union of two sets. This index is calculated by summing up the proportion of overlap between the sets and dividing this by the sum of the proportions in the union. The weighted version is particularly useful when the elements are characterized by their frequency or probability, making it suitable for applications where these attributes are critical. The detailed mathematical formulas for the weighted Jaccard index and its residual are provided in Eq.~\ref{eq:weighted jaccard} and Eq.~\ref{eq:res_weighted jaccard}:
\begin{equation}\label{eq:weighted jaccard}
J_{w}(X,Y) = \frac{\sum_{k\in{X\cap{Y}}}{min(w_{X}(k),w_{Y}(k))}}{\sum_{k\in{X\cup{Y}}}{max(w_{X}(k),w_{Y}(k))}},
\end{equation}
\begin{equation}\label{eq:res_weighted jaccard}
R_{J_{w}(X,Y)} = 1-\frac{\sum_{k\in{X\cap{Y}}}{min(w_{X}(k),w_{Y}(k))}}{\sum_{k\in{X\cup{Y}}}{max(w_{X}(k),w_{Y}(k))}},
\end{equation}
where $\sum_{k\in{X\cap{Y}}}{min(w_{X}(k),w_{Y}(k))}$ is sum of the smaller weight for each element common to both sets, and $\sum_{k\in{X\cup{Y}}}{max(w_{X}(k),w_{Y}(k))}$ determines sum of the larger weight for each elment present in either set.

\subsection{BERTScore and Residual}
While our setting involves no textual data, our portfolio similarity method can be conceptually benchmarked against the BERTScore metric adapted to structured data. The original BERTScore \cite{zhang2019bertscore} computes similarity between two texts by evaluating pairwise semantic similarity between words and aggregating them into a sentence-level score using greedy matching. In our setting, each element in a portfolio is analogous to a word in a sentence, and the similarity score between a pair of elements plays a role analogous to the BERT-based semantic similarity between words.

BERTScore computes similarity using recall, precision, and the F1 measure. Recall matches each element in the reference set 
$X$ to the most similar element in the candidate set 
$Y$ (Eq.(\ref{eq:recall})), while precision reverses this direction (Eq.(\ref{eq:precision})). Greedy matching maximizes pairwise similarity, and the F1 score combines recall and precision (Eq.(\ref{eq:f1})). To reflect the importance of rare elements in NLP, BERTScore incorporates inverse document frequency (idf) weighting. For fair comparison with our method, importance weights are instead based on the lesser proportion of each matched pair, with residuals defined in Eqs.(\ref{eq:recall_res})–(\ref{eq:f1_res}). Unlike our method, BERTScore does not update weights after each match.
\begin{subequations}
\begin{align}
  \label{eq:recall}
  R_{BERT}&=\frac{\sum^{m}_{i=1}{min(w_{X}(i),w_{Y}(j))\max_{j}S_{ij}}}{\sum^{m}_{i=1}{min(w_{X}(i),w_{Y}(j))}}, \\
  \label{eq:precision}
  P_{BERT}&=\frac{\sum^{n}_{j=1}{min(w_{Y}(j),w_{X}(i))\max_{i}S_{ji}}}{\sum^{n}_{j=1}{min(w_{Y}(j),w_{X}(i))}},\\
  \label{eq:f1}
  F_{BERT}&=2\frac{P_{BERT}\cdot{R_{BERT}}}{P_{BERT}+R_{BERT}}.
\end{align}
\end{subequations}
Residual:
\begin{subequations}
\begin{align}
  \label{eq:recall_res}
  R_{R_{BERT}}&=1-\sum_{i=1}^{m}\max{(0,w_{X}^{(m)}(i))}, \\
  \label{eq:precision_res}
  R_{P_{BERT}}&=1-\sum_{j=1}^{n}\max{(0,w_{Y}^{(n)}(j))},\\
  \label{eq:f1_res}
  R_{F_{BERT}}&=2\frac{R_{P_{BERT}}\cdot{R_{R_{BERT}}}}{R_{P_{BERT}}+R_{R_{BERT}}},
\end{align}
\end{subequations}
where $min(w_{X}(i),w_{Y}(j))$ is the smaller proportion associated with the selected pair $(i,j)$, $min(w_{Y}(j),w_{X}(i))$ is the smaller proportion associated with the selected pair $(j,i)$, and $w_{X}^{(m)}(i)$, $w_{Y}^{(n)}(j))$ correspond to the remaining proportion of element $i$ and $j$ in the sets, when the search ends after $m$ and $n$ rounds, respectively (considering zero if negative).

\subsection{Theoretical Distinctions of STRAPSim}
Before presenting the dataset, experimental setup, and comparative results, we outline key theoretical distinctions of the STRAPSim approach relative to existing similarity metrics. STRAPSim provides a flexible and interpretable framework for measuring portfolio similarity, characterized by the following core features:
\begin{enumerate}
    \item \textbf{Constituent-level Semantic Matching}: In contrast to Jaccard index variants that require exact matches, STRAPSim supports pairwise semantic similarity between portfolio constituents. It performs greedy bipartite matching by prioritizing the most similar pairs first, ensuring that high-similarity matches dominate the score. This yields more robust and granular comparisons, especially when portfolios differ in composition but serve similar roles.

    \item \textbf{Weight-aware Matching}: STRAPSim incorporates asset weights to reflect the importance of each constituent in a portfolio. Unlike Jaccard-based or token-counting metrics that treat all elements equally, STRAPSim weights similarity scores by the minimum available mass in each matched pair, providing a more realistic measure of overlap when element proportions vary significantly.

    \item \textbf{Residual-aware Dynamics}: Unlike BERTScore~\cite{zhang2019bertscore}, which treats each element's importance as static throughout the matching process, STRAPSim dynamically updates residual weights after each match. This residual-aware mechanism avoids overcounting and ensures that the total contribution of a constituent is bounded by its available share, better capturing one-to-one alignment between sets.

    \item \textbf{Comparison with Optimal Transport (OT)}: STRAPSim shares conceptual similarities with OT-based methods \cite{villani2008optimal} in that both aim to match mass across distributions. STRAPSim differs in several key aspects: it uses greedy matching rather than solving a global linear program, avoids the computational cost of solving transport maps, and naturally incorporates pairwise semantic similarities (e.g., via cosine or embedding-based distances). This makes STRAPSim more interpretable and computationally efficient, while remaining faithful to distributional alignment.

    \item \textbf{Interpretability and Transparency}: STRAPSim explicitly tracks which elements are matched and which remain unmatched, offering clear interpretability. The ability to examine residual mass after matching provides users with insight into unaligned components of the portfolio, which is particularly valuable in financial applications requiring transparency in decision-making.

    \item \textbf{Broader Applicability}: While developed for portfolio similarity, STRAPSim's design generalizes to any domain involving weighted set comparisons with soft constituent-level similarity, such as collaborative filtering, document similarity, or matching in recommendation systems.
\end{enumerate}

\section{Data Description}
This section introduces details of the public and financial datasets utilized in this study.
\subsection{Toy Datasets}
The proposed method was evaluated on various publicly available datasets \cite{Dua:2019}. Table \ref{tbl:toy-data} provides a brief summary of the toy datasets employed in this study, including datasets like Big Mac, breast cancer, and others.
\begin{table}[ht]
\centering
\fontsize{7.5}{8}\selectfont
\begin{tabularx}{\columnwidth}{lllll}
    \hline
        Data & No. of Instances & No. of Features & Source &Type\\
    \hline
    Iris & 150 & 4 & UCI &Classification\\
    Breast Cancer&699&9&UCI&Classification\\
    Big Mac & 69 & 9 & alr3 &Regression\\
    Movie Rating & 671 & 680 & Kaggle&Recommendation\\
    \hline
\end{tabularx}
\caption{Summary of Public Datasets. \label{tbl:toy-data}}
\vspace{-10mm}
\end{table}

\subsection{Corporate Bond ETFs}
The experiment was conducted on 20 corporate bond ETFs, which consists of 6870 different bonds in total, with an average value of 511 bonds per ETF. The similarity was computed using March 2024 portfolio data and return correlations based on monthly returns from February 2022 to March 2024. A random forest was first trained with OAS and yield as target variables to retrieve proximity metrics as constituent-level similarity $S_{ij}$ to support the subsequent computation of STRAPSim and others\cite{jeyapaulraj2022supervised}. The features in the training model include:
\begin{itemize}
    \item Issuer: The company that issued the bond to borrow money;
    \item Days to maturity: The number of days to bond's maturity;
    \item Industry: The industry of the bond's issuer;
    \item Bond rating: A composite rating for the bonds based on the ratings given by various credit rating agencies;
    \item Age: The number of days since the bond's issuance;
    \item Market: Market (US, EURO, China, etc.) that the bond issued in;
    \item Currency: Currency that the bond is issued in;
    \item Coupon: Annual interest paid on the bond expressed as a percentage of the bond's face value;
    \item Country: The country in which the bond is issued;
    \item flag 144a: If bond falls under "Rule 144A";
    \item Amount issued: The total issue size of the bond offering;
    \item Coupon frequency: Indicates the number of times the couppon is paid in a given year.
\end{itemize}
Categorical features (country, industry, and bond rating) were transformed using one-hot encoding. As the dataset contained no missing values, no imputation was required. The data were randomly shuffled and split into 90\% for training and 10\% for testing. We applied 5-fold cross-validation on the training set to perform hyperparameter tuning, specifically optimizing the number of trees and maximum depth of the model. Model performance was evaluated using Root Mean Squared Error (RMSE) and Mean Absolute Percentage Error (MAPE). On the training data, the model achieved an RMSE of 0.21 and a MAPE of 0.08. On the testing data, the RMSE and MAPE were 0.51 and 0.15, respectively.

\section{Experimental Set up and Results}
This section presents the experimental set up and results on the toy datasets and the corporate bonds ETFs.

\subsection{Experimental Setup}
In the toy datasets, each instance is treated as a set whose elements are weighted by different feature values. To illustrate the computation process, consider two example data points from the \textit{Iris} dataset: $X = (5.1,3.5,1.4,0.2)$ and $Y = (4.9,3.0,1.4,0.2)$ where each instance is represented as a set of four features with associated values. In the STRAPSim framework, each data point is treated as a set of constituents (features), and the similarity between instances is computed through a two-stage process.

First, a constituent-level similarity matrix is constructed by measuring pairwise similarity between features of the data, denoted as $S_{ij}$, where feature values are first scaled by their maximum to ensure comparability across dimensions. This step mirrors the common preprocessing step in regression and classification tasks that utilize cosine similarity to capture directional alignment between feature vectors. With four features in \textit{Iris}, a total of \textit{four choose two} pairwise similarities are computed as shown in Table~\ref{tbl:feature-corr}.
\begin{table}[ht]
\centering
\fontsize{7.5}{8}\selectfont
\begin{tabular}{l|llrr}
    \hline
             & sepal length          & sepal width               & \multicolumn{1}{l}{petal length} & \multicolumn{1}{l}{petal width} \\ \hline
sepal length & \multicolumn{1}{r}{1} & \multicolumn{1}{r}{0.978} & 0.948                            & 0.898                           \\
sepal width  &                       & \multicolumn{1}{r}{1}     & 0.871                            & 0.809                           \\
petal length &                       &                           & 1                                & 0.983                           \\
petal width  &                       &                           & \multicolumn{1}{l}{}             & 1                               \\ \hline
\end{tabular}
\caption{Example Cross Feature Correlation: Iris Dataset. \label{tbl:feature-corr}}
\vspace{-10mm}
\end{table}
Next, these constituent-level similarities are aggregated using the STRAPSim algorithm. The scaled feature values serve as weights, $X = (0.65,0.80,0.20,0.08)$ and $Y = (0.62,0.68,0.20,0.08)$ in this case, reflecting the relative importance of each constituent in its respective instance. STRAPSim then performs a greedy matching procedure in a recursive manner that iteratively identifies the most similar feature pairs, adjusting the weights dynamically after each match. This STRAPSim score is subsequently used within a K-nearest neighbors (KNN) framework. Specifically, for classification tasks, it serves as the similarity measure for identifying the nearest neighbors, whose labels are then used to assign the predicted class for a given instance. For regression tasks, predictions are computed as the weighted average of target values from the most similar neighbors, using STRAPSim scores as weights.

The movie rating prediction task, a typical example of recommender systems, is framed analogously. Each user is modeled as a set of movies, weighted by the user’s ratings. A constituent-level similarity matrix between movies is constructed based on their descriptions and taglines, using TF-IDF (term frequency–inverse document frequency) to convert text to numerical representations. Cosine similarity is then computed between movie pairs using the resulting term-document matrix.

Based on this matrix, user similarity is computed using the proposed STRAPSim (basket similarity) and several baseline metrics. For each user pair, the similarity between rated movies is aggregated using the constituent-level similarity scores $S_{ij}$, where user ratings serve as the item weights. For the prediction task, the rating for a new movie by a test user is estimated as a weighted average of ratings provided by 20 nearest neighbors in the training set, using user similarity as the weighting factor. All experiments are conducted using a 10-fold cross-validation scheme to ensure robustness and generalizability.

In the ETF setting, each portfolio is treated as a weighted basket of securities. Portfolio similarity is computed by optimally matching each security in the reference portfolio to the most similar security in the candidate ETF. Constituent-level similarity is derived from random forest proximity measures, as described in Ref.~\cite{jeyapaulraj2022supervised}. The same constituent-level similarity matrix is used to compute BERTScore-style metrics. For comparison, other baseline metrics—such as the Jaccard index and its weighted variants—are calculated by examining overlapping security constituents between ETF pairs.

To evaluate the alignment between similarity metrics and real market behavior, we use historical monthly return correlations as a calibration benchmark. For each ETF, we compute its return correlation with every other ETF using historical monthly returns. We then compare the resulting ranking with those produced by the various similarity metrics. The underlying assumption is that ETFs with greater constituent-level similarity should exhibit higher return correlation. To quantify this relationship, we compute the Spearman rank correlation between similarity scores and return correlations across all ETF pairs. This procedure is repeated for every ETF, and we report the average Spearman correlation along with the corresponding p-values to assess statistical significance.

\subsection{Performance Evaluation on Example Datasets}

\begin{figure}
\centering
  \includegraphics[width=0.6\columnwidth]{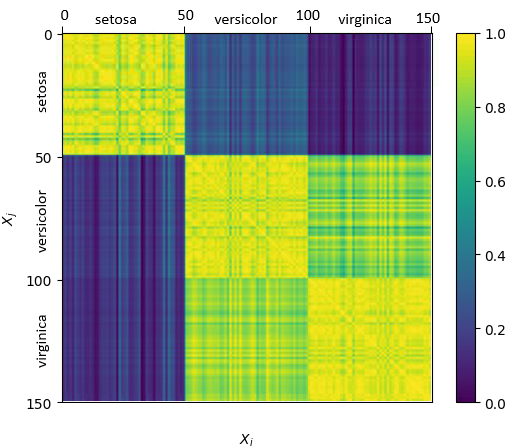}
  \caption{Heatmap of Similarity Metrics (Dataset: Iris).}\label{fig: heatmap-iris}
  \Description[]{}
  \vspace{-4mm}
\end{figure}

\begin{figure}
\centering
  \includegraphics[width=0.8\columnwidth]{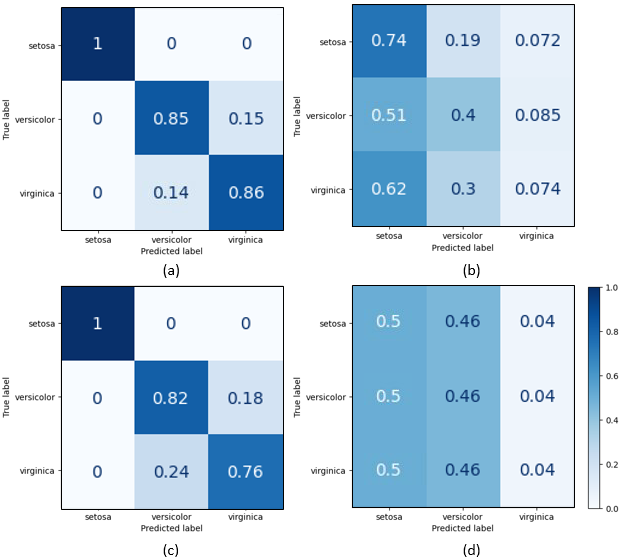}
  \caption{Confusion Matrix (Dataset: Iris). (a): STRAPSim, (b): BertScore, (c): Weighted Jaccard, (d): Jaccard Index. }
  \Description[]{}
  \label{fig: cm-iris}
    \vspace{-4mm}
\end{figure}

Table~\ref{tbl:results-toy} summarizes the results, illustrating the predictive performance of the STRAPSim method compared to other metrics. The best-performing measures are highlighted in bold. It is observed that STRAPSim consistently delivers superior results across all tasks. For classification problems, STRAPSim achieves the highest accuracy and $F_{1}$ score compared to other similarity metrics. These results are further illustrated in Figure \ref{fig: heatmap-iris} and Figure \ref{fig: cm-iris}. The heatmap demonstrates the strengths and weaknesses of relationships between data pairs evaluated by STRAPSim, with data pairs ordered by their true class labels. Notably, STRAPSim perfectly identifies the \textit{setosa} class and shows a strong capability in distinguishing between \textit{versicolor} and \textit{virginica}. This ability is further quantified by the confusion matrix, where STRAPSim produces the darkest shades in the diagonal elements, indicating superior predictive accuracy.

Even for the \textit{breast cancer} dataset, which presents imbalanced distributions, STRAPSim proves more effective at representing the pairwise closeness within the data. The higher accuracy and $F_{1}$ score demonstrate its effectiveness in distinguishing between the unbalanced classes, \textit{benign} and \textit{malignant}. By maximizing the utilization of known information, STRAPSim excels at capturing nuanced differences within data pairs and adjusting the surrounding neighborhood accordingly.

\begin{figure}
\centering
  \includegraphics[width=0.7\columnwidth]{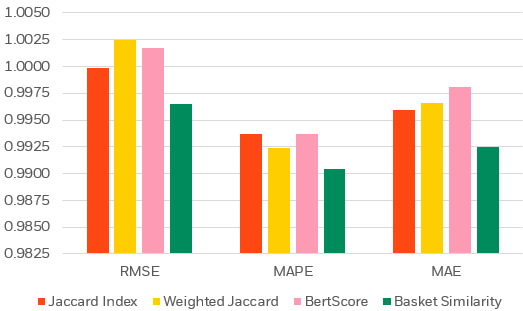}%
  \caption{Error Metrics of Movie Rating Forecasting (normalized by unweighted average).}
  \Description[]{}
  \label{fig: movie-plt}
\vspace{-6mm}
\end{figure}

\begin{figure*}
\centering
  \includegraphics[width=0.65\textwidth]{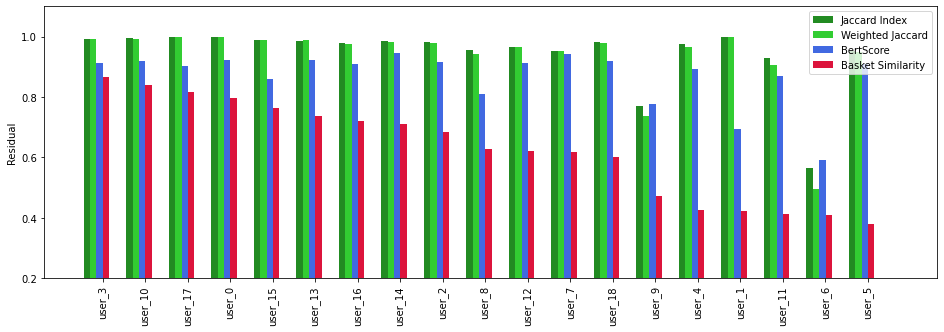}
  \caption{Residual (Dataset: Movie Rating).}
  \Description[]{}
  \label{fig: residual-movie}
    \vspace{-4mm}
\end{figure*}

For regression tasks, STRAPSim also demonstrates higher predictive accuracy, as measured by RMSE, MAPE, and MAE. Figure \ref{fig: movie-plt} illustrates the movie rating results through bar charts of the error measures, scaled by unweighted average. It is observed that STRAPSim outperforms other similarity metrics as well as the equal weighting average. This superior performance can be attributed to several advantages of STRAPSim over other similarity metrics. STRAPSim appropriately represents the data distribution by assigning weights to each element. It extends beyond exact matches by considering element-wise similarity and optimizes the utilization of known information by constantly updating the weights after each search, reflecting the evolution of the search process. Further insights are provided in Figure \ref{fig: residual-movie}, which illustrates the residual amounts left untapped by each similarity metric when computing the similarity for an individual user. STRAPSim consistently extracts the most information from each user, followed by BertScore, while the Jaccard Index tends to leave behind the most information. This enhanced capability to mine constituent-level information allows STRAPSim to more effectively learn the nuances of similarity and distance between data pairs.

\subsection{Performance Evaluation on Corporate Bond ETFs}
\begin{figure*}
\centering
  \includegraphics[width = 0.8\textwidth]{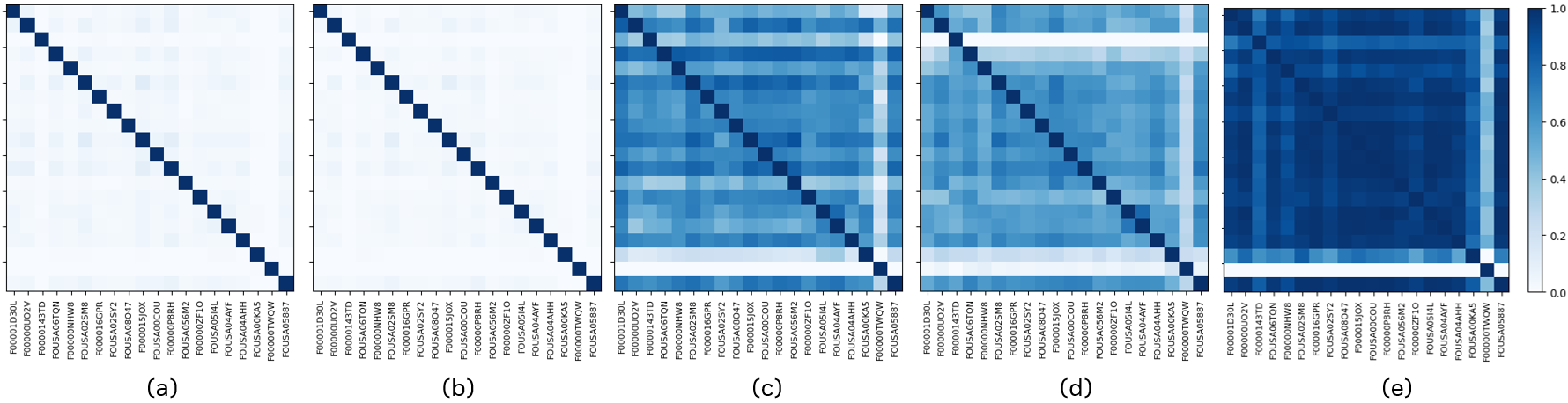}
  \caption{Heatmap of Similarity Matrices (Dataset: Corporate bonds): (a) Jaccard index, (b) weighted Jaccard, (c) BertScore, (d) STRAPSim, (e) correlation of monthly returns.}
  \Description[]{}
  \label{fig:similarity_matrices}
    \vspace{-4mm}
\end{figure*}

The STRAPSim and other similarity metrics were computed between the 20 corporate bond ETFs using information at the constituent level. Additionally, the correlation of monthly returns between the bonds was calculated to compare with the rankings derived from the similarity metrics. Figure~\ref{fig:similarity_matrices} provides illustrations of the similarity matrices between the 20 bonds. The heatmap of STRAPSim scores is found to be more align with the correlation of monthly returns, i.e., it is better at differentiating the light-shaded bonds, whereas the Jaccard index and weighted Jaccard does not deliver much information.

Table~\ref{tbl:results-fund} summarizes the statistics from the Spearman's rank correlation test. The table reports the average Spearman's rank correlation values between the ranking orders provided by the basket and other similarity metrics and the correlation of monthly returns, along with their p-values. The results indicate that the STRAPSim outperforms the other similarity metrics, achieving higher correlation coefficients and lower p-values, which demonstrates a stronger monotonic relationship with the correlation of monthly returns. The more strongly positive Spearman's correlations suggest that bonds with higher STRAPSim tend to exhibit higher correlations in their monthly returns. Furthermore, the percentage of significant samples, particularly at the 10\% significance level, reveals that STRAPSim consistently shows statistically significant rank correlations with monthly returns across all ETFs. This finding underscores the robust capability of STRAPSim to mine constituent-level information effectively and measure the distances between ETF pairs accurately. It is more powerful at identifying and ranking similar ETF pairs compared to other similarity metrics. 

\begin{table}[ht]
\centering
\fontsize{7.5}{8}\selectfont
\begin{tabularx}{\linewidth}{cccccc}
    \hline
        Dataset & Metric & Jaccard & Weighted Jaccard & BertScore & STRAPSim \\
    \hline
    {\multirow{2}{*}{{Iris}}} &Accuracy &0.33&0.86&0.41&\textbf{0.90}\\
     &F-1 Score&0.49&0.86&0.52&\textbf{0.90}\\
     \hline
     Breast &Accuracy &0.39&0.67&0.50&\textbf{0.68}\\
     Cancer&F-1 Score&0.56&0.70&0.56&\textbf{0.72}\\
     \hline
    {\multirow{3}{*}{{Big Mac}}} &RMSE &30.91&21.37&37.94&\textbf{20.11}\\
     &MAPE(\%)&84.89&26.25&113.27&\textbf{25.21}\\
     &MAE&22.36&11.35&29.21&\textbf{10.73}\\
     \hline
     Movie &RMSE &0.91&0.91&0.91&\textbf{0.90}\\
     Rating&MAPE(\%)&23.35&23.32&23.35&\textbf{23.27}\\
     &MAE&0.71&0.7129&0.71&\textbf{0.71}\\
     \hline
\end{tabularx}
\caption{Summary of Results on Toy Datasets. \label{tbl:results-toy}}
\vspace{-8mm}
\end{table}


\begin{table}[ht]
\fontsize{7.5}{8}\selectfont
\centering

\begin{tabularx}{\linewidth}{XXXXX}
    \hline
          Similarity Metrics        & Average Coefficient & Average P-value & \% of significant samples ($\alpha$=5\%) & \% of significant samples ($\alpha$=10\%) \\
    \hline
Jaccard     & 0.5864              & 0.0791          & 80                                        & 90                                         \\
weighted Jaccard  & 0.5791              & 0.0592          & 90                                        & 90                                         \\
BertScore         & 0.5865              & 0.0548          & \textbf{95}                                        & 95                                         \\
STRAPSim & \textbf{0.6783 }             &\textbf{0.0081}           & \textbf{95}                                        & \textbf{100}                                       \\
    \hline
\end{tabularx}

\caption{Statistics of Spearman’s Rank Correlation. \label{tbl:results-fund}}
\vspace{-6mm}
\end{table}

\section{Conclusion and Outlook}

Portfolio similarity underpins key financial workflows including fund benchmarking, trade execution, and liquidity sourcing, particularly in bond ETF markets where illiquidity and heterogeneous holdings challenge standard metrics. We propose STRAPSim, a constituent-similarity-aware, weight-sensitive, and residual-aware similarity measure tailored for comparing real-world portfolios.

Unlike traditional metrics such as the Jaccard index \cite{jaccard1901etude,costa2021further} or proximity-based methods \cite{jeyapaulraj2022supervised, desai2023quantifying}, STRAPSim performs greedy matching between partially overlapping assets, dynamically adjusts weights, and aggregates scores using a residual-aware mechanism. Drawing inspiration from BERTScore \cite{zhang2019bertscore} in NLP, STRAPSim adapts its logic and expands the concept further to structured financial data while maintaining interpretability and robustness in low-frequency environments.

Empirical results across public benchmarks and ETF holdings demonstrate that STRAPSim consistently outperforms baselines in classification, regression, and correlation-based evaluation. Notably, it achieves the highest Spearman correlation with return-based ETF similarity, affirming its utility as a proxy for economic co-movement.

STRAPSim also supports portfolio trading by enabling accurate matching between bespoke baskets and benchmark ETFs, improving pricing transparency and operational scalability. Future work will focus on integrating STRAPSim into live trading and construction pipelines and extending it to broader asset classes.
\section{Acknowledgement}
The views expressed here are those of the authors alone and not of BlackRock, Inc.

\bibliographystyle{ACM-Reference-Format}
\bibliography{sample-base}
\end{document}